**Effect of defect-induced cooling on graphene hot-electron bolometers.**

Abdel El Fatimy[a]*†, Peize Han[a], Nicholas Quirk[a], Luke St. Marie[a], Matthew T. Dejarld[b], Rachael L. Myers-Ward[b], Kevin Daniels[c], Shojan Pavunny[b], D. Kurt Gaskill[b], Yigit Aytac[c], Thomas E. Murphy[c] and Paola Barbara[a]*.

[a]Department of Physics, Georgetown University, Washington, DC 20057, USA.

[b] U.S. Naval Research Laboratory, Washington, DC 20375, USA,

[c] Institute for Research in Electronics and Applied Physics, University of Maryland, College Park, MD 20742, USA.

## Abstract

At high phonon temperature, defect-mediated electron-phonon collisions (supercollisions) in graphene allow for larger energy transfer and faster cooling of hot electrons than the normal, momentum-conserving electron-phonon collisions. Disorder also affects the heat flow between electrons and phonons at very low phonon temperature, where the phonon wavelength exceeds the mean free path. In both cases, the cooling rate is predicted to exhibit a characteristic cubic power law dependence on the electron temperature, markedly different from the $T^4$ dependence predicted for pristine graphene. The impact of defect-induced cooling on the performance of optoelectronic devices is still largely unexplored. Here we study the cooling mechanism of hot-electron bolometers based on epitaxial graphene quantum dots where the defect density can be controlled with the fabrication process. The devices with high defect density exhibit the cubic power law. Defect-induced cooling yields a slower increase of the thermal conductance with increasing temperature, thereby greatly enhancing the device responsivity compared to devices with lower defect density and operating with normal-collision cooling.

## *Corresponding Authors

*Abdel El Fatimy, a.elfatimy@gmail.com; *Paola Barbara, Paola.Barbara@georgetown.edu

## †Present Addresses

† Ecole Centrale Casablanca, Bouskoura, Ville Verte, 27182, Casablanca, Morocco.

## 1. Introduction

Understanding the effect of disorder on the relaxation dynamics of charge carriers is crucial for the operation of most graphene devices. When electrons absorb energy, they quickly thermalize via electron-electron collisions within a few tens of femtoseconds [1-3] and via emission of optical phonons ($\hbar\omega_{op}$ > 200 meV) within a few hundreds of femtoseconds [4, 5], reaching an electron temperature $T_e$ that can be substantially higher than the temperature of the graphene lattice. Dissipation of energy via acoustic phonons yields much longer cooling times, from tens to hundreds of picoseconds [6, 7]. This is because momentum conservation and the limited Fermi surface of a two-dimensional material like graphene severely constraint the maximum amount of energy that can be dissipated in a collision process. The most energetic acoustic phonons are emitted by electrons that are backscattered. For an electron with energy E, this maximum phonon energy is given by $\Delta E_{ph} = 2Ev_s/v_F$, where $v_F$ is the Fermi velocity and $v_s$ is the sound velocity in graphene. Since $v_F \sim 50\ v_s$, this energy loss is a few percent of the electronic energy and therefore it takes many collisions for electrons to cool down. The maximum energy transfer to acoustic phonons for electrons at the Fermi energy defines the Bloch-Grüneisen temperature, $T_{BG} = 2E_F v_s/(v_F K_B)$, a characteristic temperature that plays a role similar to the Debye temperature in three-dimensional conductors. At temperatures below $T_{BG}$, the normal-collision cooling is characterized by the power law $P = A\Sigma\ (T_e^4 - T_0^4)$, where $\Sigma = \frac{\pi^2 D^2 |E_F| K_B{}^4}{15\rho_M \hbar^5 v_F^3 v_s^3}$ is a coupling constant,

D is the deformation potential of graphene, $\rho_M$ its charge density, A is the graphene area, $T_0$ is the temperature of the graphene lattice and P is the power absorbed by the graphene under optical or electrical (Joule) pumping [8, 9]. (Here we are assuming that the system is in a steady state, $C_e\left(\frac{dT_e}{dt}\right) = P - Q = 0,$ where $C_e$ is the electronic heat capacity and the absorbed power P balances the energy loss $Q$ due to electron cooling.) At higher temperatures, $T_{BG} << T_0$, $T_e << E_F/K_B$, the power is predicted to depend linearly on temperature, $P \propto (T_e - T_0)$ [9].

In the presence of defects, the momentum conservation constraints described above are relaxed, because defect-assisted collisions enable emission of phonons with higher energy and momentum than normal collisions [7]. This supercollision-cooling regime is characterized by faster cooling times and by a cubic power law in steady state, $P = A\,\Sigma_2\,(T_e^3 - T_0^3)$, where $\Sigma_2 = \frac{\zeta(3) D^2 |E_F| K_B{}^3}{\pi^2 \rho_M \hbar^4 v_F^3 v_s^2 l_{mfp}}$, $l_{mfp}$ is the mean free path and the Riemann zeta function $\zeta(3) \approx 1.2$ [7]. Previous work on CVD grown graphene [10] and exfoliated graphene [11] measured this cubic power law at temperatures higher than $T_{BG}$, where, in the presence of defects, the supercollisions are predicted to dominate over normal collisions [7]. Other work used pump-probe experiments on exfoliated graphene with the defect density systematically increased by exposure to near-infrared femtosecond pulses and showed that samples with higher defect density had faster cooling times [12]. For lower temperatures, $T_x < T_0$, $T_e < T_{BG}$, the cooling occurs via normal collision, with the dependence $P \propto (T_e^4 - T_0^4)$ [10] down to a crossover temperature $T_x = \frac{30\,\hbar\,v_s\,\zeta(3)}{\pi^2 K_B l_{mfp}}$ [13, 14]. When the electron and phonon temperatures are lowered below $T_x$, the cooling power is predicted to regain a cubic law dependence $P = A\,\Sigma_3\,(T_e^3 - T_0^3)$, with $\Sigma_3 = 2\Sigma_2$ [13, 14].

The importance of charge carrier cooling in graphene goes well beyond understanding the basic physics of two-dimensional conductors because it impacts any application of this material for bolometry and photodetection. However, to date the effect of defect-assisted cooling on the performance of graphene detectors is still unclear.

Here we study high-performance bolometers based on quantum dots of epitaxial graphene on silicon carbide. We show that the defect density in the graphene and the cooling mechanism can be controlled by the fabrication process. We find that devices based on defective graphene exhibit defect-assisted cooling and yield higher responsivity over a wider dynamic range than devices made with graphene having low defect density and operating under normal collision cooling.

## 2. Results and Discussion

We recently showed that quantum dots patterned from epitaxial graphene on SiC exhibit a very strong dependence of the electrical resistance on temperature, higher than 100 MΩ $K^{-1}$, yielding extraordinary values of bolometric responsivity, larger than $10^9$ V $W^{-1}$ [15, 16]. The strong temperature dependence of the resistance is caused by the quantum confinement gap of the dots and depends on the dot diameter. (The orientation of the devices with respect to the steps between adjacent crystal planes on the SiC substrates also affects the temperature dependence of the resistance, because they affect the current flow through the device [15]). Figure 1(a) shows the resistance vs. temperature curves for three samples.

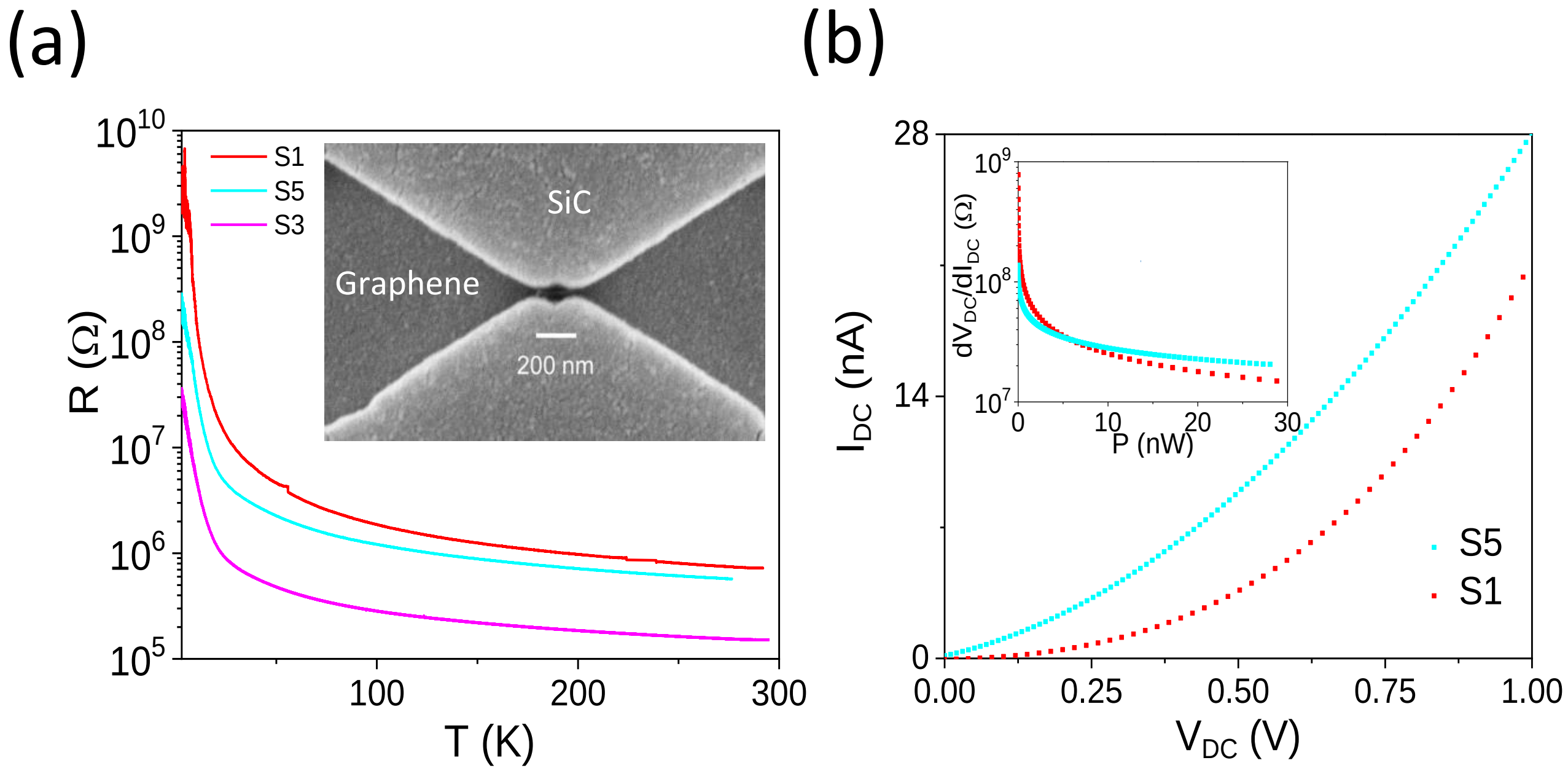

**Figure 1.** (a) Resistance vs. temperature dependence for three quantum dot devices with diameters of 30 nm (S1), and 200 nm (S3 and S5). The data for device S1 are reproduced from ref. [15]. (b) Current vs. voltage (IV) characteristics of two devices, at $T_0$ = 3K. The inset shows the differential resistance $dV_{DC}/dI_{DC}$ as a function of Joule power ($P = I_{DC} \cdot V_{DC}$) obtained from the IV curves.

Typical current-voltage (IV) characteristics of the quantum dot devices are shown in Figure 1(b). The IV curves are non-linear because, as the bias voltage increases, the device temperature increases due to Joule heating and the resistance decreases, corresponding to the resistance vs. temperature (R(T)) curves in Figure 1(a). The inset in Figure 1(b) shows the differential resistance as a function of Joule power, computed from the IV curves of the devices. By combining the data of the differential resistance vs. Joule power with the R(T) data, we readily obtain the dependence of the electron temperature on Joule power that can be used to investigate the cooling mechanism of the quantum dot devices.

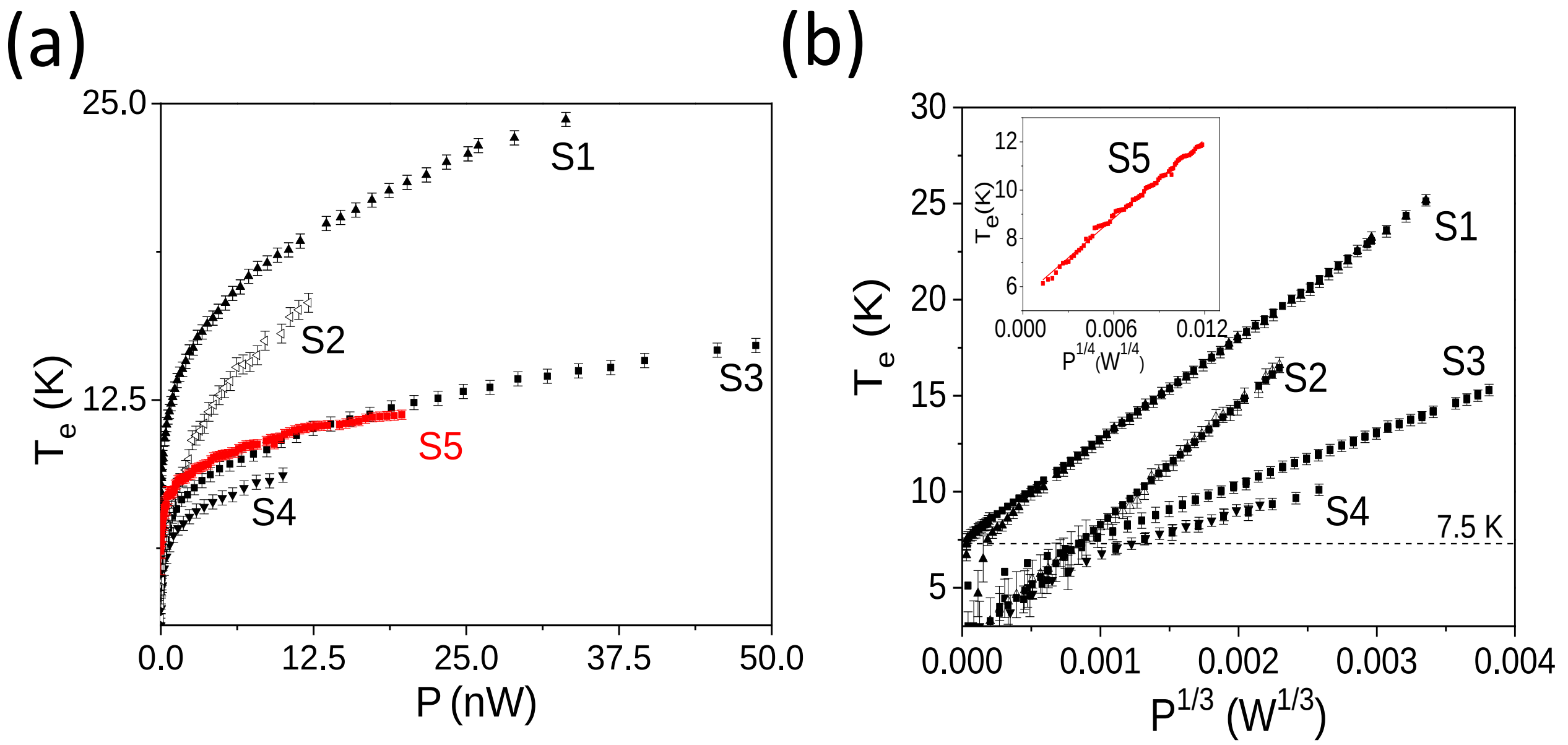

**Figure 2.** (a) Electron temperature vs. Joule power for the sample patterned with overdosed PMMA (S5) and for the four samples patterned with a thin metallic layer (S1, S2, S3 and S4). (b) Data from (a) showing that $T_e$ scales linearly with $P^{1/3}$ for samples S1, S2, S3 and S4 for $T_e > 7.5$ K. Inset: For sample S5, $T_e$ scales linearly with $P^{1/4}$.

Most of the devices studied here and in our previous work [15, 16] were fabricated using a thin (30 nm of Au or Pd) metallic layer as a mask to pattern the quantum dots [15, 17]. The metallic layer protected the graphene from contamination with photoresist and was removed using aqua regia as the last fabrication step [17]. We also fabricated a test device using a different fabrication process, where the graphene was patterned by using crosslinked PMMA instead of the metallic layer. The data for five different samples are shown in Figure 2 (a), with the red symbols showing the data for the test sample patterned with PMMA and the black ones showing the data for four samples patterned using the thin metallic layer. To test the cubic power law, we plot these same data as $T_e$ vs. $P^{1/3}$ in Figure 2 (b). We find that, for $T_e > 7.5$ K (dotted line in Figure 2 (b)), all the samples fabricated using the metal layer as a mask to pattern the quantum dots show a very good agreement with the cubic power law, whereas the sample fabricated using the crosslinked PMMA, sample S5, shows a good agreement with the prediction for normal collision cooling, $P \propto T_e^4$ (see inset of Figure 2(b)).

All the samples were fabricated using graphene epitaxially grown on SiC, but the different power laws clearly indicate that different fabrication procedures yield samples with different cooling mechanisms. The fabrication procedure can also affect the doping of the graphene. As-grown graphene on SiC typically exhibits n-type doping, with charge density of about $n \sim 10^{12}$ $cm^{-2}$, corresponding to $T_{BG} \sim 70$ K. This is consistent with the power law dependence expected for normal collisions in the low-temperature regime and measured for the sample fabricated using the overdosed PMMA. Conversely, the doping level of the graphene samples fabricated using the thin metal layer is expected to be quite different, due to the aqua regia treatment used to remove the metal layer. The aqua regia is a p-dopant for graphene [17], therefore the charge carrier density and value of $T_{BG}$ can be quite different for the samples fabricated with the thin metal layer.

Although all the samples operated roughly in the same temperature range, $T_0$, $T_e < 30$ K, the cubic temperature dependence of the samples fabricated using the metal layer mask indicates that 1) the cooling mechanism is dominated by defects and 2) the samples could either be in the low-temperature regime, with $T_0$, $T_e < T_x$, or in the high temperature regime, with $T_0$, $T_e > T_{BG}$. Since $T_{BG}$ depends on the graphene charge density via $E_F$, it is important to investigate how the different fabrication processes may affect the defect density and the doping of graphene.

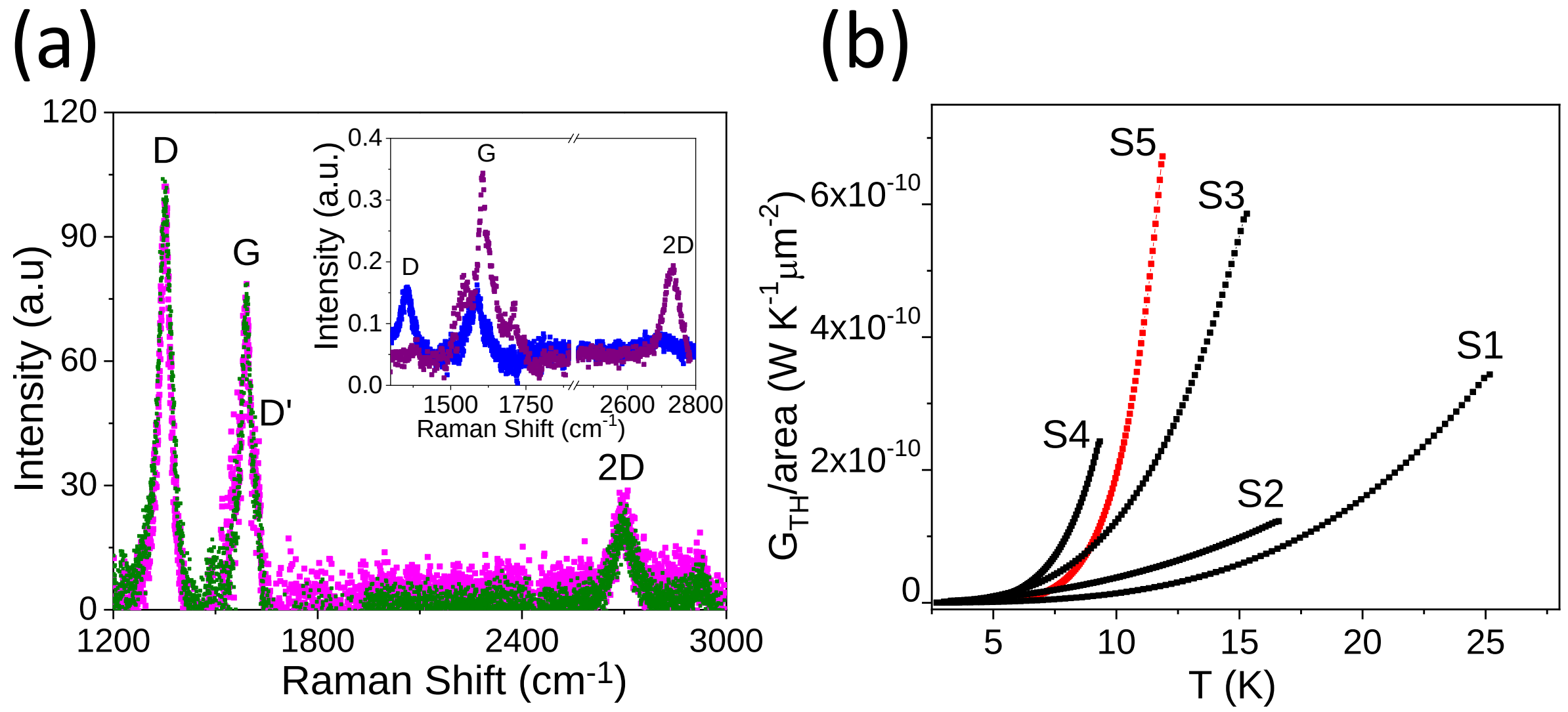

**Figure 3.** (a) Raman spectra of samples S3 (pink) and S4 (green) after subtracting the SiC substrate. The inset shows the Raman spectra of sample S5'for the part protected with PMMA (purple) layer and the part processed with Au sputtering and aqua regia (blue). All spectra are taken at $E_L$ = 2.33 eV ($\lambda_L$ = 532 nm). (b) Thermal conductance per unit area for all the samples, obtained from the data in Figure 2.

Figure 3 (a) shows the Raman spectra of samples S3 and S4 fabricated using the sputtered thin metallic layer, after subtracting the Raman spectrum of the SiC substrate. The Raman spectra were measured with a Horiba LabRAM HR Evolution, with a 532 nm laser (Ventus 532 from Laser Quantum). The inset shows the spectrum of another test sample, S5', made of graphene epitaxially grown on SiC, with a small area of graphene protected by a layer of overdosed PMMA (purple), using the same procedure we used for sample S5. A thin metallic layer was subsequently deposited on the whole sample S5' and then removed with aqua regia. The Raman spectrum after this procedure is also shown in the inset (blue). Sample S5' was fabricated to unambiguously test the effect of the metal deposition on the quality of the exposed graphene and compare it to the graphene protected by the PMMA layer, using graphene epitaxially grown on the same SiC substrate. The samples S3 and S4 clearly show the presence of a large defect (D) peak at 1350 cm$^{-1}$, higher than the G peak ($I_D/I_G$ = 1.3). By contrast, the graphene covered with PMMA on sample S5’ shows no measurable D peak. On the same sample S5’, the part of graphene that was not protected with PMMA shows a substantially reduced intensity of the 2D peak and a large D peak ($I_D/I_G = 0.9$), similar to sample S3 and S4. For epitaxial graphene on SiC there is a contribution to the D peak from the buffer layer [18], but this is considerably smaller than the D peak found for the metal coated samples, thus indicating considerable reduction of sp2 symmetry caused by the presence of defects. For the PMMA-coated sample, the buffer layer contribution is similar in magnitude and shape to that reported by Fromm et al. [18] and observed for untreated pristine samples. Hence, there is little or no D peak for the PMMA coated sample. The data clearly indicate that the metal deposition introduced defects that were not present in the pristine or protected samples.

To further test the effects of the fabrication process on the defect density and doping, we divided a single sample of graphene epitaxially grown on SiC into three sections SA, SB, and SC, and processed each section differently. The graphene in section SA had a thin metallic layer sputtered on it and then removed with aqua regia, in the same way as during the fabrication of graphene quantum dots for samples S1, S2, S3, and S4. The graphene in section SB was patterned with overdosed PMMA, in the same manner as the fabrication of graphene quantum dots for sample S5. The graphene in section SC was removed by oxygen plasma etching, exposing the SiC substrate. We then measured the Raman spectrum, mobility, and doping of the graphene in each section.

The Raman spectra of sections SA and SB are shown in Figure 4(a), with the SiC background subtracted. Similar to the samples discussed in Figure 3(a), the sample SA shows a substantial increase of the defect peak. The defects are induced in the graphene by the metal deposition and they depend on the deposition method, as described in the Supplementary Information. Values of the mobility and the doping level also differ substantially between the two regions, as shown in Table 1. The region treated with metal sputtering shows low mobility, about 51 $cm^2\ V^{-1}s^{-1}$, and very high hole doping, $5.6\times10^{12}$ $cm^{-2}$, whereas the region protected with PMMA has mobility about five times higher, 260 $cm^2\ V^{-1}s^{-1}$, and electron doping with concentration typical for epitaxial graphene grown on SiC. It is clear that the sputtering and aqua regia process strongly dope the graphene with holes: it completely compensates the electron doping of the graphene epitaxially grown on SiC and further dopes it with holes, reaching charge density values comparable to the electron doping of the other region. The mean free paths for the two regions, $l_{mfp} = \mu\ (\pi n)^{1/2}\ (\hbar/e)$, where $n$ is the charge carrier density and μ is the mobility, are also listed in Table 1.

| Fabrication process | Charge density ($cm^{-2}$) | μ ($cm^2\ V^{-1}\ s^{-1}$) | $l_{mfp}$ (nm) |
|---|---|---|---|
| SA (Pd sputtering and aqua regia) | $+ 5.6\times10^{12}$ | 51 | 1.3 |
| SB (PMMA protection) | $- 3.4\times10^{12}$ | 260 | 5.3 |

**Table 1.** Characteristics of epitaxial graphene on SiC treated with different fabrication processes. The measurements were done at room temperature.

Using the values in Table 1, we can estimate the characteristic temperatures $T_{BG}$ and $T_x$. Due to the similar values of charge carrier density, the values of $T_{BG}$ are similar for the graphene treated with metal deposition and aqua regia and the graphene protected by PMMA (128 K and 100 K, respectively). The values of $T_x$ are very different because of the different mean free paths, yielding $T_x$ = 42 K for the graphene treated with Pd sputtering and aqua regia and a lower value, about 10 K, for the graphene protected with PMMA. This means that even though all the bolometers operate in the same temperature range, below 30K, those treated with metal sputtering and aqua regia operate in the low-temperature regime, with $T_0$, $T_e < T_x$, where the cubic power dependence holds, whereas the bolometer with the graphene protected by PMMA operates at temperatures very close to or higher than the crossover temperature $T_x$, where the temperature dependence for normal collisions dominates.

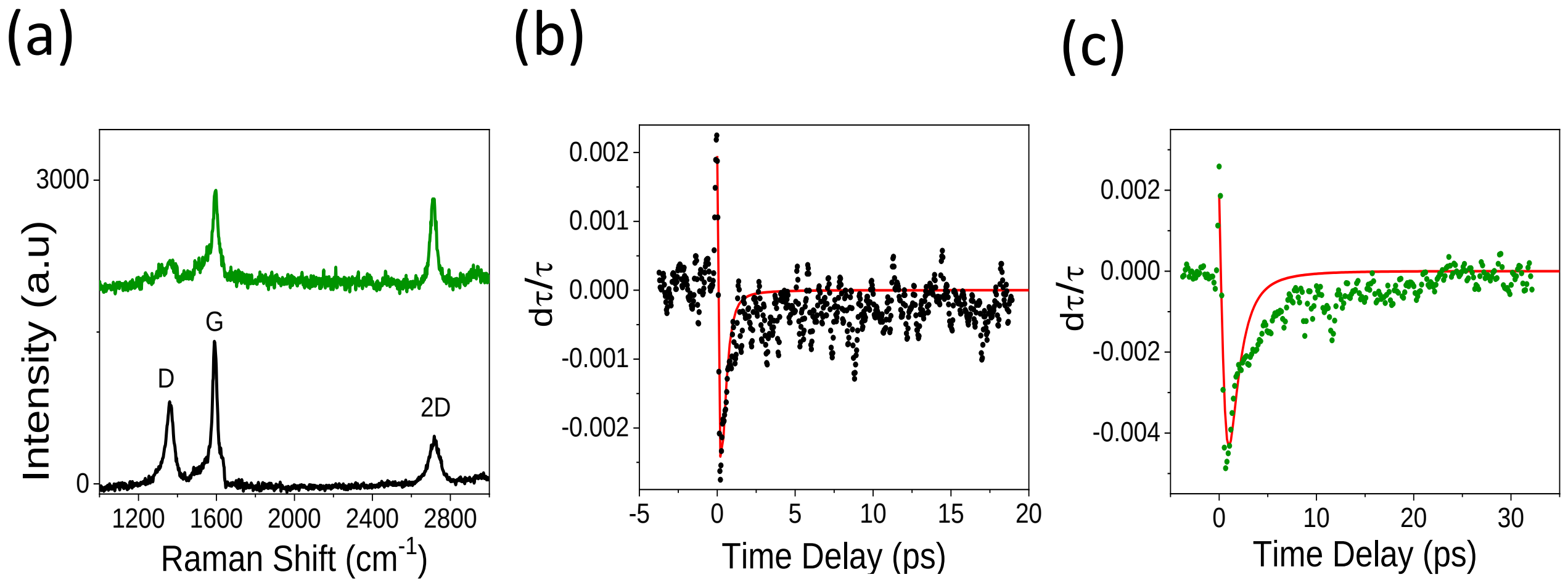

**Figure 4**: (a) Raman spectra of graphene on SiC substrates for two separate regions: region SA treated with Pd sputtering subsequently removed with aqua regia (black) and region SB with graphene protected with PMMA (green). The curves are shifted vertically for clarity. All spectra are taken at $E_L$ = 2.33 eV ($\lambda_L$ = 532 nm). (b) and (c) Time-resolved differential-transmission measurements for Pd sputtered (SA) and PMMA covered (SB) graphene on SiC, respectively, measured at room temperature. The red curves show fits to the data, using the model discussed in the text.

From the slopes of the $T_e$ vs. $P^{1/3}$ plots, ranging from 0.18 to 6 W $K^{-3}m^{-2}$, we can extract the deformation potential D and find that it varies from 3 eV to 18 eV for the different samples treated with metal sputtering and aqua regia. These values are within the range found in other studies [11, 19-21]. We also extracted the deformation potential from the slope of the $T_e$ vs. $P^{1/4}$

of the PMMA-protected bolometer and found a larger value of D, about 40 eV, but still within the range of values found from previous work [10].

Having established that the quality of the graphene and the cooling mechanism can be controlled with the fabrication process, it is now important to understand which type of graphene and cooling mechanism will yield the best bolometric performance. Supercollision cooling yields faster devices, but its effect on one of the most important figures of merit, the bolometric responsivity, has not yet been explored. The responsivity is defined as the change of voltage $\Delta V_{DC}$ across the device caused by the incident light divided by the absorbed power, $r = \Delta V_{DC}/\Delta P = I_{DC}(\Delta R/\Delta P) = (I_{DC}/G_{TH})(\Delta R/\Delta T)$, where $G_{TH}$ is the thermal conductance, $\Delta R$ is the change in resistance caused by a temperature increase $\Delta T$ and $\Delta V_{DC}$ is measured at a constant current $I_{DC}$. As discussed above and in our previous work [15, 16], the temperature dependence of the quantum dot resistance is mainly determined by the quantum dot diameter, but the thermal conductance and its temperature dependence will strongly depend on the graphene quality and the graphene cooling mechanism. We can extract the thermal conductance directly from the data in Figure 2 using $G_{TH} = dP/dT_e$. Figure 3 (b) shows the thermal conductance as a function of temperature for all the samples discussed above. Although the values of thermal conductance of all the samples are comparable at low temperature, it is clear that the thermal conductance increases more slowly for the samples with supercollision cooling compared to the temperature dependence of the thermal conductance of the PMMA-protected sample undergoing normal collision cooling, as expected from the different exponents of the power law for the two different mechanisms. Since high responsivity requires lower thermal conductance, it is clear that the samples undergoing supercollision cooling will yield higher responsivity in a wider dynamic range.

So far we discussed the cooling properties of electrons in steady state, where the power absorbed from the incident radiation balances the energy lost via cooling. Now we focus on the regime where $C_e\left(\frac{dT_e}{dt}\right) = P - Q \neq 0,$ to see how the defects induced by the different fabrication processes affect the cooling dynamics of electrons in graphene. We performed pump-probe measurements of the samples SA and SB at room temperature. The measurements were performed in the mid-IR spectral range, with wavelengths of the pump pulse set to 1.4 μm and the probe at 5.5 μm (see Figure S1 in the Supplementary Information). The pump and probe were approximately 100 fs in duration, with radii ($e^{-1}$ of the intensity) of ≃100 μm and ≃35 μm, respectively, both much smaller than the size of the SA and SB regions, which ensures that the probe beam samples a spatially uniform region of photoexcitation.

The time dependence of the differential transmission for sections SA and SB is shown in Figure 4 (b) and (c), respectively. The differential transmission shows positive values right after the pulse, when $T_e$ is highest, then decreases sharply to negative values and reaches a minimum before increasing towards vanishing values. This non-monotonic time-dependence of the differential transmission can be used to extract information on the time dependence of the electron temperature. In the pump-probe measurements, the pump beam is optically chopped, while the probe beam is synchronously detected, thereby measuring the fractional change in transmission $\Delta\tau/\tau_0 = (\tau-\tau_0)/\tau_0$ as a function of the relative time difference between the pump and probe pulse, where $\tau_0$ represents the optical transmission of the (room-temperature) graphene film in the absence of a pump pulse and $\tau$ is the time-dependent transmission under optical illumination. The mid-infrared optical transmission through the graphene sample depends implicitly on the electron temperature, through the relation $\tau = \frac{4Y_1 Y_2}{|Y_1 + Y_2 + \sigma(\omega, T_e)|^2}$ [22], where $Y_1$ and $Y_2$ represent the admittances of the incident and substrate regions respectively, and $\sigma(\omega, T_e)$ is the complex conductivity of the graphene sheet, which is related to the temperature through the Kubo model [23]. The initial electron temperature, immediately following the absorption of the pump pulse, can be approximated as $T_{e,peak} = \left[T_0^2 + 2\frac{\eta\, P_{pump}}{\alpha\ \pi\ w^2}\right]^{1/2}$. Here $\alpha = \frac{C_e}{AT} = \frac{2\pi k_B^2 E_F}{3\ \hbar^2 v_F^2}$ is the heat capacity coefficient, $\eta$ is the fractional absorption in the graphene film, $P_{pump}$ is the pump power and $w$ is the width of the optical beam at the focus. For the fluence and beam diameter used in the experiments, we estimate $T_{e,peak}$ = 10,000 K. Assuming that supercollisions are the dominant cooling mechanism (we note that the pump-probe measurements are performed at room temperature, above $T_{BG}$ for both SA and SB), the electron temperature relaxes to the lattice according to the dynamical equation: $\alpha T_e(t)\frac{dT_e(t)}{dt} + \Sigma_2[T_e(t)^3 - T_0^3] = 0$. Figures 4(b) and (c) show the best fits to the data for samples SA and SB, using the parameters shown in Table 2.

| | SB (PMMA Covered) | SA (Pd sputtering and aqua regia) |
|---|---|---|
| $n_0\,(cm^{-2})$ | $-3.4 \times 10^{12}$ | $11 \times 10^{12}$ |
| $\mu\ (cm^2/Vs)$ | 260 | 90 |
| $l$ (nm) | 5.4 | 3.5 |
| $\alpha\ (J/m^2K^2)$ | $1.2 \times 10^{-9}$ | $2.2 \times 10^{-9}$ |
| $\Sigma_2 (W/m^2K^3)$ | 0.28 | 1.3 |

**Table 2**: Parameters used for the temperature model fits on time resolved measurements in Figure 4(b) and (c).

The mobility and charge carrier density values yielding the best fits are in very good agreement with the parameters independently measured by Hall measurements on the PMMA-covered sample (SB), whereas they are off by about a factor of two for the sample that was treated with Pd sputtering (SA). We note that the pump probe measurements were performed about one month before the transport measurements and the sample properties might have slightly deteriorated for the unprotected sample SA, due to sample handling and its exposure to ambient conditions. Nevertheless, the pump probe measurements confirm that the cooling coefficient $\Sigma_2$ is higher for the sample treated with Pd sputtering, leading to a faster response time and consistent with a shorter mean free path and higher defect concentration [7, 12]. Moreover, the values of obtained $\Sigma_2$ from the pump-probe measurements are well within the range of the values obtained from the linear fits of $T_e$ vs. $P^{1/3}$ from the transport measurements of the quantum dot bolometer samples in Figure 2(b).

## 3. Conclusions

In summary, we studied the cooling mechanism of graphene epitaxially grown on SiC. We show that the fabrication process, including the deposition of thin metallic layers and exposure to aqua regia, substantially affects the defect density and the cooling mechanism. We find that the combination of faster response time and lower thermal conductance in a wide range of power and temperature make defective graphene the best material for bolometric applications.

## Acknowledgements

This work was supported by the US Office of Naval Research (N00014-16-1-2674) and the NSF (ECCS-1610953). Research at NRL is supported by the Office of Naval Research. Matt and Shojan are grateful for support through the ASEE Postdoctoral Program.

**Appendix A. Supplementary Data**

Supplementary data associated with this article can be found in the online version.

## Appendix A: Supplementary Data

**Pump Probe Measurements:**

The optical pump-probe measurements were performed using a regeneratively amplified Ti-sapphire laser (Coherent Libra) and optical parametric oscillator, which produces signal and idler pulses of approximately 100 fs duration at wavelengths of 1400 nm and 1880 nm, respectively. The signal at 1440 nm was used to optically stimulate the graphene samples under study. The signal and idler were further mixed in an external $AgGaS_2$ crystal to produce mid-infrared pulses at 5500 nm, which were used to probe the hot-carrier response of the graphene. The spectral content of the pump and the probe signal is shown in Figure S1, as obtained from FTIR measurements.

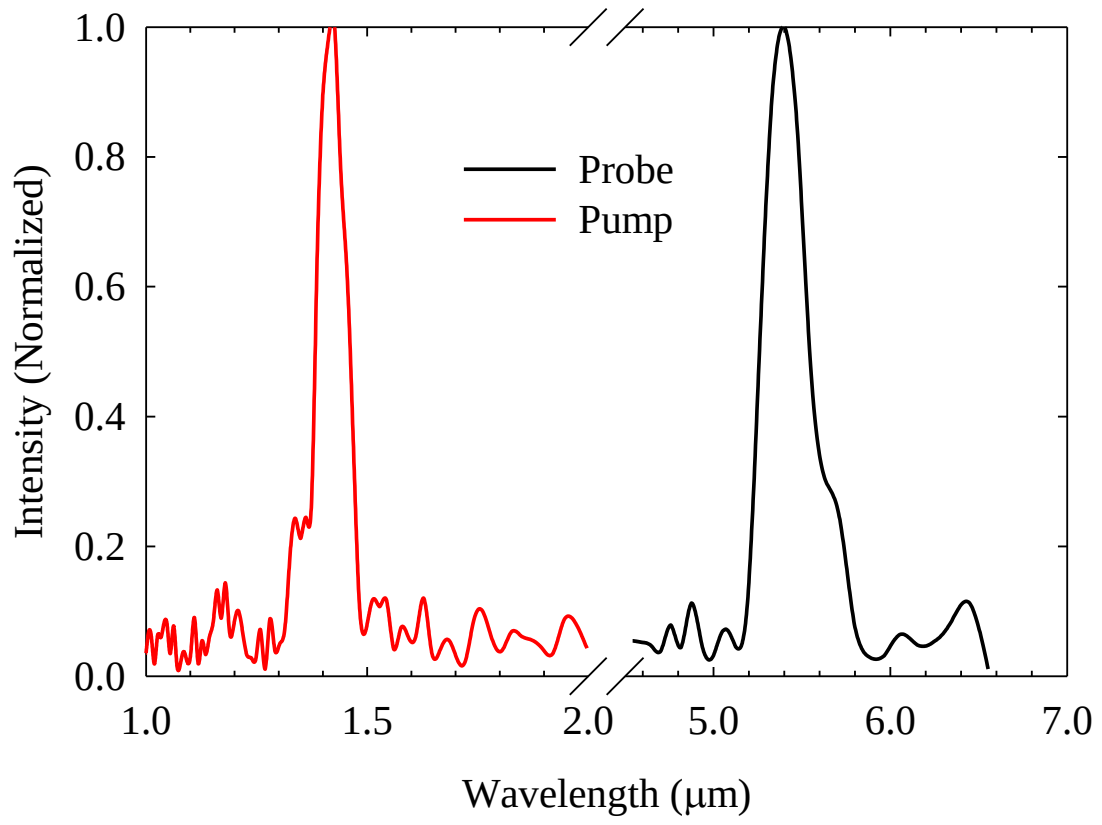

**Figure S1**: Measured spectrum of pump and probe beams, as measured with FTIR.

The pump and probe were co-polarized and co-focused onto the graphene samples under study, while the relative timing between them was adjusted using a delay stage. The table below outlines the experimental parameters of the 1 kHz co-polarized pump-probe measurements.

| | Pump | Probe |
|---|---|---|
| Wavelength ($\mu m$) | 1.4 | 5.5 |
| Energy ($meV$) | 885 | 225 |
| Power ($\mu W$) | 400 | <10 |
| Spot Size ($\mu m$) | 100 | 40 |

**Table S1**: Parameters used for pump probe experiments.

## Peak Electron Temperature

The evolution of electron temperature, $T_e$, in graphene can be described as:

$$\alpha T_e(t)\frac{dT(t)}{dt} + \beta_1\,[T_e(t) - T_0] + \Sigma_2[T_e(t)^3 - T_0^3] + \Sigma[T_e(t)^4 - T_0^4] = \eta\, I(t)$$

where $I(t)$ represents the intensity of the optical pump pulse, $\eta$ is the fractional absorption in the graphene and the other parameters were introduced in the main manuscript.

Because the pump pulse is very short in duration compared to the relevant cooling timescales, the intensity $I(t)$ may be approximated as a delta function, $I(t) = F_0\,\delta(t)$ with $F_0 = \frac{P_{pump}}{\pi\, w^2 R}$, where $P_{pump}$ is the pump power and $w$ is the width of optical beam at the focus.

Integrating the differential equation from $t = 0^-$ to $t = 0^+$, the initial electron temperature is found to be: $T_{e,peak} = \left[T_0^2 + 2\frac{\eta\, P_{pump}}{\alpha\,\pi\, w^2}\right]^{1/2}$.

where the $\alpha$ is the heat capacity coefficient, $\alpha = \frac{C_e}{AT}$ and it is a function of the Fermi energy, as shown in the parameter summary table.

| | | |
|---|---|---|
| Heat Capacity coefficient | $\alpha = \frac{2\pi k_B^2 E_F}{3\ \hbar^2 v_F^2}$ | $\frac{J}{m^2 K^2}$ |
| Ordinary-cooling coefficient | $\beta_1 = \frac{V_D^2 E_F^4 k_B}{2\ \pi\rho\ \hbar^5 v_F^6}$ | $\frac{W}{m^2 K}$ |
| Super-cooling coefficient | $\Sigma_3 = \frac{\zeta(3) V_D^2 E_F k_B^3}{\pi^2 \rho \hbar^4 v_F^3 s^2 l}$ | $\frac{W}{m^2 K^3}$ |
| Fermi energy, carrier density | $n = \frac{E_F^2}{\pi \hbar^2 v_F^2}$ | $J, \quad \frac{1}{m^2}$ |
| Fermi velocity | $v_F = 10^6$ | $m/s$ |
| Chemical Potential | $\mu_c = \sqrt{n_0 \pi \hbar^2 v_F^2}$ | $J$ |
| Areal Density | $\rho = 7.6 \times 10^{-7}$ | $\frac{kg}{m^2}$ |
| Acoustic velocity | $s = 2.1 \times 10^4$ | $\frac{m}{s}$ |
| Deformation potential | $V_D = 10$ | $eV$ |
| Carrier mobility | $\mu = e\frac{v_F}{\sqrt{\pi n}\Gamma\hbar}$ | $\frac{m^2}{Vs}$ |

**Table S2:** Graphene parameters.

## Optical properties:

A linearly polarized wave is normally incident on a two-dimensional conductive sheet, the transmission, reflection and absorption can be calculated using a simple transmission line model, where the incident and substrate regions are modeled with impedances $Z_1 = Z_0/\sqrt{\epsilon_1}$, and $Z_2 = Z_0/\sqrt{\epsilon_2}$. The frequency-dependent conductance of the graphene sheet is $\sigma(\omega)$. Using the notation $Z_{1(2)} = 1/Y_{1(2)}$), the reflection, transmission and absorption are found to be

$$r(\omega) = \left|\frac{Y_1 - Y_2 - \sigma(\omega)}{Y_1 + Y_2 + \sigma(\omega)}\right|^2 \quad \text{(Reflection)}$$

$$\tau(\omega) = \frac{4Y_1 Y_2}{|Y_1 + Y_2 + \sigma(\omega)|^2} \text{ (Transmission)}$$

$$a(\omega) = 1 - r(\omega) - \tau(\omega) = \frac{4Y_1 \, Re\{\sigma(\omega)\}}{|Y_1 + Y_2 + \sigma(\omega)|^2} \text{ (Absorption)}$$

The conductivity can be separated in intra- and inter-band contributions,

$$\sigma(\omega) = \sigma_{intra}(\omega) + \sigma_{inter}(\omega)$$

$$\sigma_{intra}(\omega) = 2\, i\, e^2\, k_B T_e \frac{\ln\left[2 \text{ Cosh}\left(\frac{|\mu_c|}{2k_B T_e}\right)\right]}{\pi\, \hbar^2(\omega + i\,\Gamma)},$$

$$\sigma_{inter}(\omega) = \frac{e^2}{4\hbar}\left[\mathrm{H}(\hbar\omega - 2\mu_c) - \frac{i}{\pi} \ln\left(\frac{|\hbar\omega - 2\mu_c|}{|\hbar\omega + 2\mu_c|}\right)\right]$$

where, H is the step function that determines the cut-off edge of the interband absorption in graphene. The carrier scattering rate Γ also varies with the electron temperature [1] and it is described as:

$$\Gamma(T_e) = \Gamma_0\left(1 + \frac{\pi^2 k_B^2 T_e^2}{6\, E_F^2}\right) + \frac{E_F V_D^2 k_B T_e}{4\hbar^3 v_F^2 \rho s^2}.$$

**<u>Data Analysis:</u>**

The optical excitation pulse's temporal width is 150 fs and used to generate the non-equilibrium state in graphene. From the previous sections of this document, we have estimated the initial electron temperature of ~10,000 K. For the measurements taken at room temperature and the conditions of $T_{e-peak} \gg T_0$, the evolution of temperature in graphene, from peak to equilibrium, can be determined as follows:

$$\alpha T_e(t)\frac{dT_e(t)}{dt} + \beta_3[T_e(t)^3 - T_0^3] = 0.$$

Here we assume that the cooling is dominated by the cubic term due to supercollisions.

The solution $T_e(t)$ is plotted in Figure S2 (the parameters used in the calculations are listed in the caption).

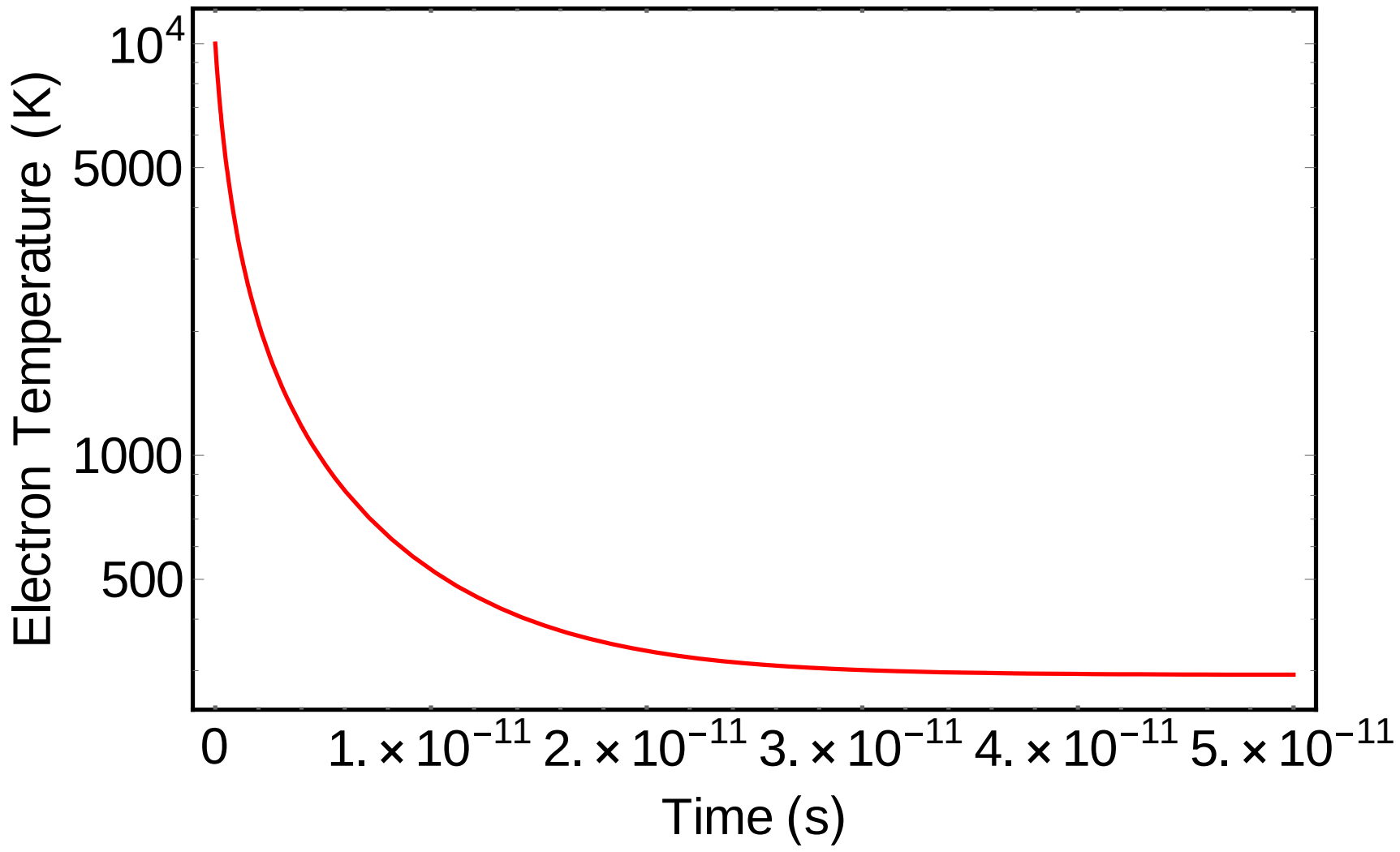

**Figure S2:** Time dependence of the electron temperature calculated from the model discussed in the text with parameters $\alpha = 2 \times 10^{-9} J/m^2K^2$ , $\beta_3 = 0.4\, W/m^2K^3$, $\mu = 100\ cm^2/Vs$, $n_0 = 10 \times 10^{12} cm^{-2}$ and $T_{e,peak} = 10{,}000\ K$.

The next task is to calculate the differential-transmission (i.e., fractional change in transmission) as a function of electron temperature, using the Kubo model of conductivity with the transmission line model of reflection and transmission. The fractional change in temperature caused by hot electrons is

$$\frac{\Delta\tau}{\tau} = \frac{\tau - \tau_0}{\tau_0}$$

where $\tau_0$ represents the transmission in the absence of optical pumping and $\tau$ is the instantaneous transmission caused by hot carriers. For the same parameters listed above for the $T_e(t)$ calculations, the equivalent Fermi energy is estimated to be 368 meV. The fractional change $\Delta\tau/\tau$ as a function of electron temperature is shown below.

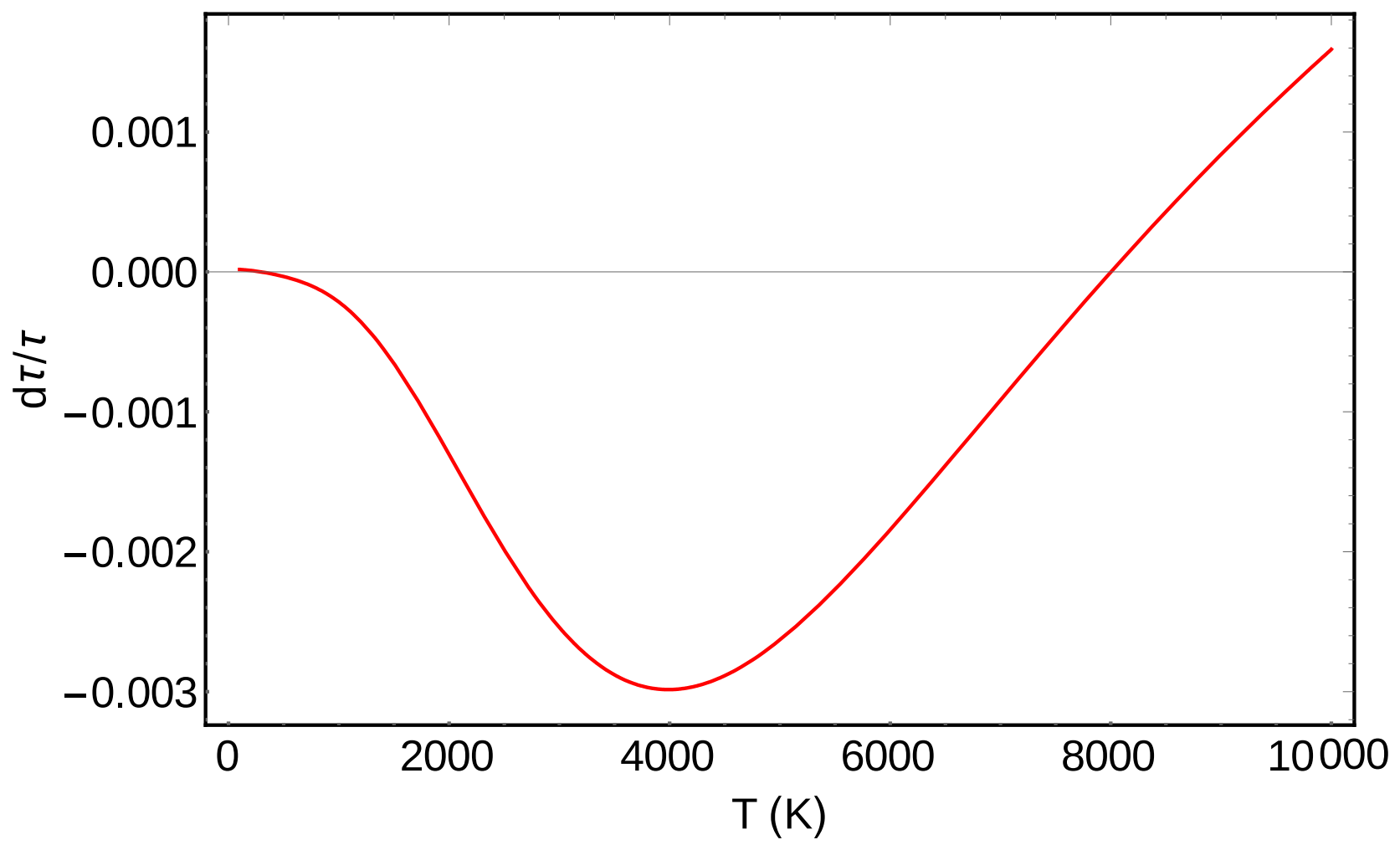

**Figure S3:** Temperature dependence of the differential transmission.

We note that at high carrier temperatures, the transmission increases. Then the electron temperature cools down to the 4000 K range, where the maximum change in negative differential transmission occurs. The minimum in the negative signal is determined by the carrier concentration and the mobility values.

Next, we insert the time-dependent electron temperature $T_e(t)$ into $\Delta\tau/\tau$ to obtain the time-resolved differential transmission values,

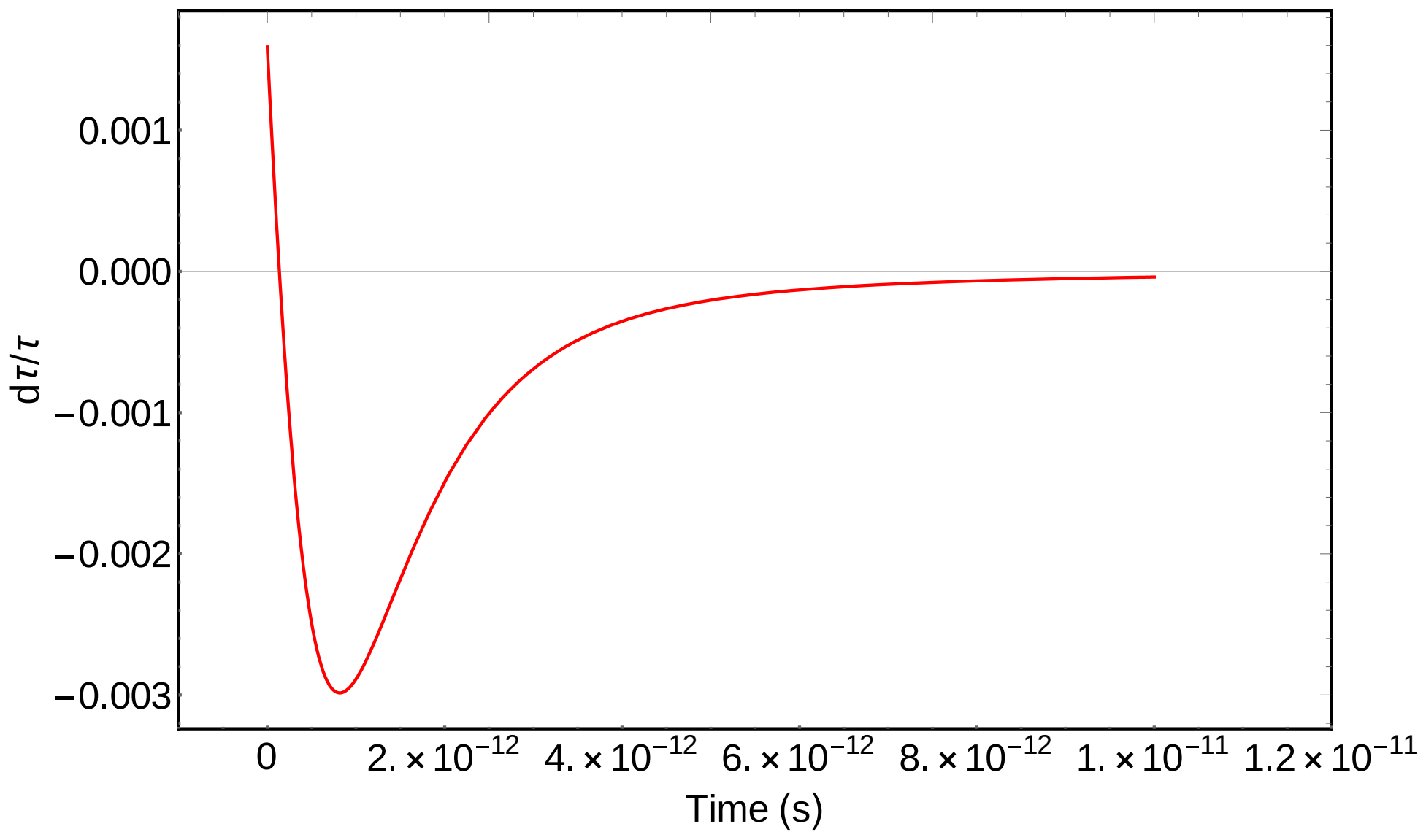

**Figure S4.** Time dependence of the differential transmission obtained by combining the data in Fig. S2 and S3.

At t = 0, the electrons begin at their peak temperature value, $T_{e,peak}$, where the differential transmission is positive. As the electrons cool, the differential transmission changes sign from positive to negative, reaching a minimum fractional change of –0.3%, before recovering to zero with a time-scale in the picosecond range.

**Metal Deposition:**

The samples S1, S2, S3, S4, S5, S5', and SA were all sputtered with palladium using argon-plasma magnetron sputtering at 60 W. To test the effects of different metal deposition methods, we fabricated additional samples with CVD-grown graphene transferred on Si substrates that were capped with 300 nm of $SiO_2$. Some graphene samples were coated with sputtered Pd, sputtered Au, or had Au deposited by thermal evaporation. In all cases, the metal was subsequently removed by aqua regia, using the same procedure that was used for the epitaxial graphene on SiC samples described in the manuscript. The Raman spectra for an uncoated (as-grown) graphene sample, a graphene sample with Au sputtered at 3 W then removed with aqua regia, and a graphene sample with Au deposited by thermal evaporation then removed with aqua regia are shown below.

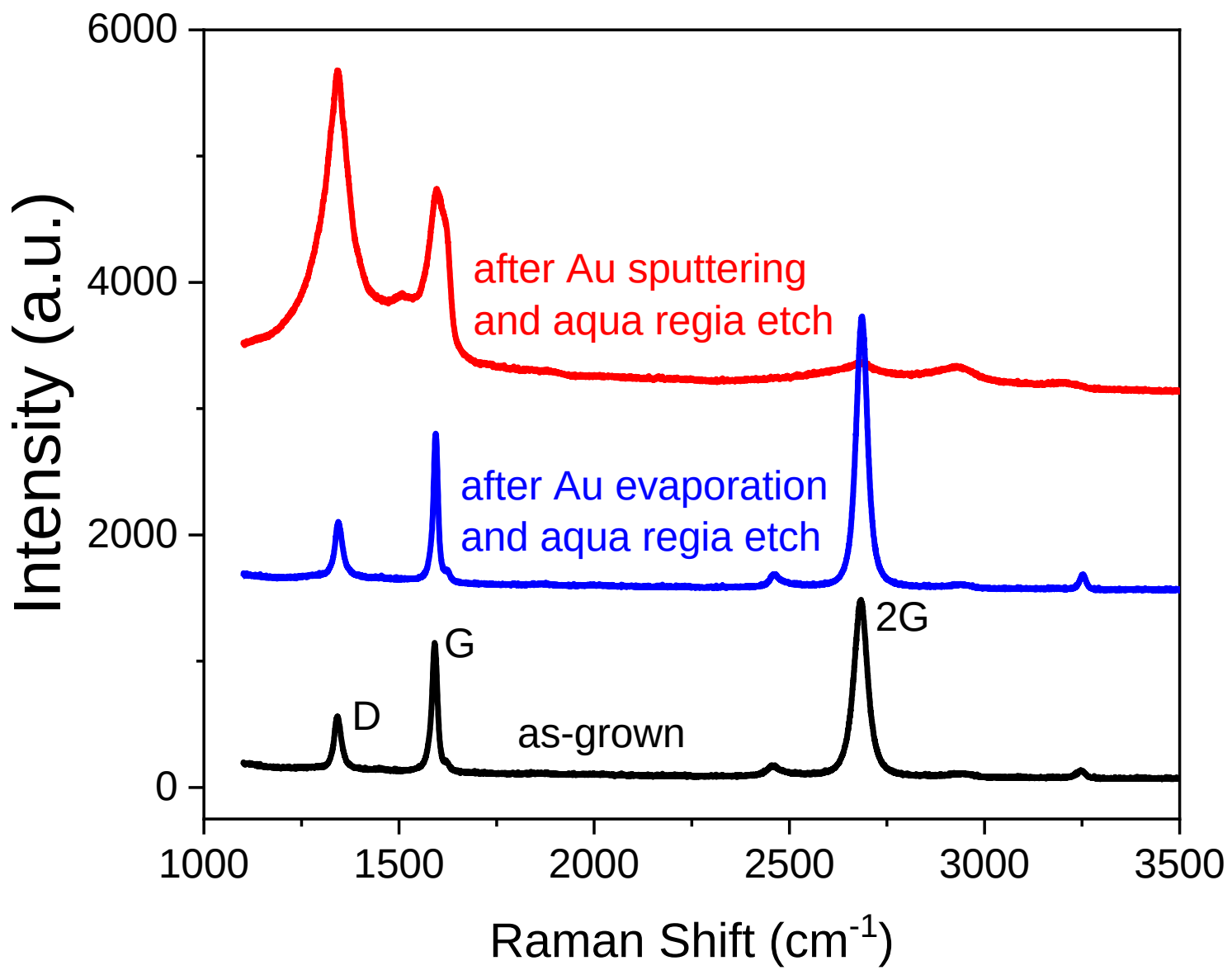

**Figure S6**: Raman spectra of CVD-grown graphene transferred on Si/$SiO_2$ substrate for an unprocessed sample (black), a sample coated with evaporated Au subsequently removed by aqua regia (blue) and a sample coated with sputtered Au subsequently removed with aqua regia (red). Both Au layers were 30-nm thick. The curves are shifted vertically for clarity.

From the Raman spectra, we see that sputtering has a much larger effect on the graphene than does thermal evaporation. Sputtering causes the D peak to increase greatly while the 2D peak is severely diminished. Evaporation and aqua regia cause much less change to the peaks.